\documentclass[aps,floatfix,twocolumn,nofootinbib]{revtex4}

\usepackage{graphics}
\newcommand{\bea}{\begin{eqnarray}}
\newcommand{\eea}{\end{eqnarray}}
\newcommand{\be}{\begin{equation}}
\newcommand{\ee}{\end{equation}}

\def\be{\begin{eqnarray}}
\def\ee{\end{eqnarray}}
\def\bd{\begin{displaymath}}
\def\ed{\end{displaymath}}

\def\etal{{\em et al. }}
\def\ADNDT{{\em At. Data. Nucl. Data. Tables }}

\def\NP{{\em Nucl. Phys. A }}
\def\PR{{\em Phys. Rev. C }}

\def\PRL{{\em Phys. Rev. Lett. }}

\def\jpg{{\em J. Phys. G: Nucl. Part. Phys. }}
\def\EPJ{{\em Eur. Phys. J. A }}

\begin{document}
\title{\bf{Effects of $NN$ potentials on $p$ nuclides in the A$\sim$100-120 region}}

\author{C. Lahiri, S. K. Biswal and S. K. Patra\\
Institute of Physics, Sachivalaya Marg, Bhubaneswar 751005, India}


\begin{abstract}
Microscopic optical potentials for low energy proton reactions have been obtained by 
folding density dependent M3Y interaction derived from nuclear matter calculation with densities from mean field approach to study
astrophysically important proton rich nuclei in 
mass 100-120 region. We compare S factors for low-energy $(p,\gamma)$ reactions with available experimental data and further
calculate astrophysical reaction rates for $(p,\gamma)$ and $(p,n)$ reactions. Again we choose some nonlinear R3Y interactions
from RMF calculation and folded them with corresponding RMF densities to reproduce experimental S factor values in this mass region.
Finally the effect of nonlinearity on our result is discussed.

\end{abstract}

\maketitle

\section{Introduction}
In nature 35 nuclei, commonly termed as $p$ nuclei,  can be found on the proton-rich side of the nuclear landscape ranging 
between $^{74}$Se to $^{196}$Hg. As they are neutron deficit, the astrophysical reactions involved in the synthesis of these 
elements do not correspond to the slow($s$) or  fast($r$) neutron capture processes. It mainly includes
reactions such as proton capture, charge exchange and photo-disintegration.  One can find a detailed study related to 
the $p$  process
in standard textbooks 
[for example, Illiadis\cite{ili}] and reviews\cite{prev}.

The natural abundances for $p$ nuclei are very low in
the order of 0.01\% to 1\%. In general, the calculation of isotopic abundances 
require a network calculation typically involving 2000 nuclei
and approximately 20000 reaction and decay channels and 
one  major problem with this $p$ network is that most of the nuclei involved in the reaction network are very shortly lived.
As a consequence, it is  
very difficult to track the $p$ process nucleosynthesis network experimentally.  
However, recent radioactive ion beam facilities are giving new prospects, still we are far away from measuring astrophysical
reaction rates for the main reactions involved in the $p$ process. Thus, one often has to depend on theoretical models
to study these reactions. 
These type of calculations acutely exploit the Hauser-Feshbach formalism where the optical model potential, in a local or a
global form, is a key ingredient. Rauscher \etal substantially calculated astrophysical reaction rates and cross 
sections in a global approach\cite{rate}. They further made a comment that the statistical model 
calculations may be improved by using locally tuned parameterization. 

In this manuscript, we perform a fully microscopic calculation. The framework is
based on microscopic optical model utilizing the theoretical  density profile of a nucleus. In presence of a stable target,
electron scattering experiment can be performed to avail nuclear charge density distribution data. However, in absence of a stable target, 
theory remains a sole guide to describe the density. Therefore, in this work we employ
relativistic mean field (RMF) approach to extract the density information of a nucleus. This has the advantage 
of extending it to  unknown mass regions. In some earlier works\cite{gg,clah1,clah2,clah3,saumi}, this method has been used
to study low energy proton reactions in the A $\sim$ 55-100 region. In a recent work\cite{dipti}, similar method has been used in 
110-125 mass region. 

The nonlinearity in the scalar field\cite{rein,sahu} in a RMF theory 
has been proved very successful in reproducing various observables 
like nuclear ground state including nuclear matter properties and the 
surface phenomena like proton radioactivity etc. In the present work, 
we intend to study the effect of  microscopic optical potentials obtained from nonlinear $NN$ interactions also in 
addition to the conventional linear $NN$ interactions in the A$\sim$ 100-120 region. In this work, we concentrate mainly on 
the region relevant to the $p$ network and therefore mainly proton rich and stability region of the nuclear landscape is our 
main concern. 

\section{Technique}
The RMF approach has successfully explained various features of stable 
and exotic nuclei like ground state binding energy, radius, deformation, 
spin-orbit splitting, neutron halo etc
\cite{walecka74,pring96,bender03,soret86,miller72}.
The RMF theory is nothing but the relativistic generalization of the 
non-relativistic effective theory like Skyrme and Gogny approach. This theory 
does the same job, what the non-relativistic theory can do, with an additional guarantee that
 it works in a better way in high density region\cite{compare}. In this manuscript we have used the RMF formalism in both direct and indirect 
way. Directly we have calculated the nuclear density, which is an essential 
quantity to calculate the optical potential. Indirectly we used  RMF Lagrangian to derive $NN$ 
interactions also along with the phenomenologically availed $NN$ interaction model. 
Here we have used different types of $NN$ interactions, namely 
the density dependent M3Y interaction (DDM3Y) and 
nonlinear R3Y interactions(NR3Y). The concept of the NR3Y was originally developed from 
basic idea of the RMF formalism\cite{bhuyan12} and will be discussed later in this section. 

In order to calculate the nuclear density, different forms of Lagrangian densities can be used from RMF approach. In this manuscript, 
the chosen form of the interaction Lagrangian density is given by
\begin{eqnarray}
{\cal L}_{int}&=&\bar{{\psi}}\bigg[{g_\sigma}{\phi}-\bigg({g_\omega}{V_\mu}+\frac{g_\rho}{2}
{\tau}.{\mathbf b_\mu}+\frac{e}{2}(1+\tau_3){A_\mu}\bigg)\gamma_{\mu}\bigg]{\psi}
\nonumber\\
&-&\frac{g_2}{3}{\phi^3}-\frac{g_3}{4}\phi^4+\frac{\xi}{4}
({V_\mu}{V^\mu})^2\nonumber\\
&+&\Lambda(\mathbf b_\mu. \mathbf b^\mu)(V_\mu V^\mu).
\end{eqnarray}
Here $m_\sigma$, $m_\rho$, $m_\omega$ are the masses of the various mesons 
like sigma, rho, and omega respectively, where as $g_\sigma$, $g_\rho$, 
and $g_\omega$ are the corresponding coupling constants given in Table \ref{nl3}.  
The coupling constants for nonlinear terms of sigma are $g_2$ and $g_3$, that for omega meson is given by $\xi$ and $\Lambda$ denotes the 
cross coupling strength between rho and omega meson.




For example, in case of FSUGold parameter set\cite{fsu}, one can see that, apart from the usual
nucleon-meson interaction terms, it contains two additional nonlinear meson-meson 
self interaction  terms including isoscalar meson self interactions, and 
mixed isoscalar-isovector coupling, whose main virtue is the softening of 
both the Equation of State (EOS) of symmetric matter and symmetry energy. 
As a result, the new parameterization becomes more effective in reproducing 
quite a few nuclear collective modes, namely the breathing modes in $^{90}$Zr and $^{208}$Pb, 
and the isovector giant dipole resonance in $^{208}$Pb\cite{fsu}. 

Again there are many other
parameter sets as well as the Lagrangian densities in RMF which are different 
from each other in various ways like inclusion of new interaction or different 
value of masses and coupling constants of the meson etc. For the comparison and 
better analysis we have included different parameter sets (NL3, TM1) as it is a matter of great concern 
 to check their credibility in astrophysical prediction. 
 Therefore, for the  astrophysical calculations we have used nuclear densities from different sets of parameters like NL3 and TM1
and folded them with corresponding $NN$ interactions respectively. In case of DDM3Y interaction, which is not obtained 
 from the RMF theory, we folded it with RMF density from FSUGold. This FSUGold folded DDM3Y interaction have been used
 in earlier works\cite{gg,clah1,clah2,clah3,saumi} and successfully reproduced some astrophysically important cross sections 
 and reaction rates
 in A $\sim$ 55-100 region. 
 Therefore, we used this potential in A $\sim$ 100-120 region as an extension of earlier works.


Typically, a microscopic optical model potential is obtained by folding an  
effective interaction, derived either from the nuclear matter calculation, in the 
local density approximation, {\em i.e.} by substituting the nuclear matter 
density with the density distribution of the finite nucleus(for example DDM3Y), or directly by folding different R3Y interactions 
using different sets of parameters 
from RMF with corresponding density distributions. 
 The folded potential therefore takes the form

\begin{equation}
 V(E,\vec R)=\int\rho(\vec {r'})v_{eff}(r,\rho,E) \vec{dr'},
\end{equation}
with $\vec r=\vec {r'}-\vec R$ in fm. These effective interactions ($v_{eff}(r,\rho,E)$) are described below in more details.


The  density dependent M3Y (DDM3Y)  
interaction\cite{ddm3y} is obtained from a finite range energy independent G-matrix 
elements
of the Reid potential by adding a zero range energy dependent pseudo-potential 
and introducing a density dependent factor.
The interaction is given by

\begin{equation}
 v_{eff}(r)=t^{M3Y}(r,E)g(\rho).                 
\end{equation} 
Here $v_{eff}(r)$ is a function of r, $\rho$ and $E$, where $E$ is incident energy and $\rho$, the nuclear density. 
The $t^{M3Y}$ interaction is defines as
\begin{equation}
t^{M3Y}=7999\frac{e^{-4r}}{4r}-2134\frac{e^{-2.5r}}{2.5r}+J_{00}(E)\delta(r)
\end{equation} 
with the zero range pseudo potential $J_{00}(E)$ given by, 
\begin{equation}
J_{00}(E)=-276\left( 1-0.005\frac{E}{A}\right) {\rm MeV} fm^{3}\end{equation} 
and $g(\rho)$ is the density dependent  factor expressed as,
\begin{equation}
g(\rho)=C(1-b\rho^{2/3})\end{equation} 
with $C=2.07$ and $b=1.624$ $fm^2$ \cite{ddm3y}.


Here the zero range pseudo potential $J_{00}(E)$ is given in equation (5).

 In \cite{sahu}, Sahu \etal introduced a simple form of nonlinear self-coupling of the scalar meson field and suggested 
a new $NN$ potential in relativistic mean field theory (RMFT) analogous to the M3Y interaction. Rather than using usual
phenomenological prescriptions, the authors derived the microscopic $NN$ interaction from the RMF theory Lagrangian. 
Starting with the nonlinear relativistic mean field Lagrangian density  for a nucleon-meson many-body system they solved the 
nuclear system under the mean-field approximation using the  Lagrangian and obtained the field equations for the nucleons and mesons.
It is necessary here to mention that the authors \cite{sahu} had taken the nonlinear part of the scalar meson $\sigma$ proportional
to $\sigma^3$ and $\sigma^4$ in account and used those terms in the opposite sign to the source term. Finally for a normal nuclear medium
the resultant effective nucleon-nucleon interaction, obtained from the summation of the scalar and vector parts of the single
meson fields takes the form\footnote{There is a typographical mistake in the expression of $v_{eff}$ in Sahu \etal \cite{sahu} and the corrected 
 form is given in this manuscript.}

\begin{eqnarray}
 v_{eff}(r) = \frac{g_{\omega}^2}{4\pi}\frac{e^{-m_\omega r}}{r}+\frac{g_{\rho}^2}{4\pi}\frac{e^{-m_\rho r}}{r}
 -\frac{g_{\sigma}^2}{4\pi}\frac{e^{-m_\sigma r}}{r}\\\nonumber
 +\frac{g_2^2}{4\pi}{r}{e^{-2m_\sigma r}}+\frac{g_3^2}{4\pi}\frac{e^{-3m_\sigma r}}{r}-\frac{\xi^2}{4\pi}\frac{e^{-3m_\omega r}}{r}\\\nonumber
 +J_{00}(E)\delta(r).
\end{eqnarray}
 \begin{table}

\begin{center}
\caption{Model parameters for the Lagrangian FSUGold\cite{fsu}, NL3\cite{lala} and TM1\cite{soga94}.\label{nl3}}
\vspace{.5cm}
\begin{tabular}{|c|c|c|c|}
\hline
{}&FSUGold&NL3&TM1\\
\hline
$M$ (MeV)&939&939&938\\

$m_\sigma$ (MeV) &491.500&508.194&511.198\\

$m_\omega$ (MeV)&782.500&782.501&783.000\\

$m_\rho$ (MeV)&763.000&763.000&770.000\\

$g_\sigma$&10.592&10.2170&10.0290\\

$g_\omega$&14.298 &12.8680&12.6140\\

$g_\rho$&11.767&4.4740&4.6320\\

$g_2$ (fm$^{-1}$)&-4.2380&-10.4310&-7.2330\\

$g_3$&-49.8050&-28.8850&0.6180\\

$\xi$&2.0460&{-}&71.3070\\

$\Lambda$&0.0300&{-}&{-}\\


\hline

\end{tabular}
\end{center}
\end{table}

 Using NL3 parameters from Table \ref{nl3}, equation (7) becomes\cite{sahu}
 
 \begin{eqnarray}
 v_{eff}(r) = 10395\frac{e^{-3.97r}}{4r}+1257\frac{e^{-3.87r}}{4r}
 -6554\frac{e^{-2.58r}}{4r}\\\nonumber
 +6830{r}\frac{e^{-5.15r}}{4}+52384\frac{e^{-7.73r}}{4r}+J_{00}(E)\delta(r).
\end{eqnarray}

The authors\cite{sahu} denoted this $NN$ interaction potential as NR3Y(NL3). 
Further, putting parameter sets from TM1 (Table \ref{nl3}),
one can obtain $V_{eff}$ for  NR3Y(TM1).






Since the DDM3Y folded potential described above do not include any spin-orbit term, the
spin-orbit potential from the Scheerbaum prescription\cite{SO} has been coupled with the
phenomenological complex potential depths $\lambda_{vso}$ and $\lambda_{wso}$ 
has been incorporated. The spin-orbit potential is given by
\begin{equation} U^{so}_{n(p)}(r)=(\lambda_{vso} +i\lambda_{wso})
\frac{1}{r}\frac{d}{dr}(\frac{2}{3}\rho_{p(n)}+\frac{1}{3}\rho_{n(p)}).
\end{equation}
The depths are functions of energy, given by 
\begin{displaymath}\lambda_{vso}=130\exp(-0.013E)+40\end{displaymath}
and 
\begin{displaymath}\lambda_{wso}=-0.2(E-20)\end{displaymath}
where $E$ is in MeV. These standard values have been used in the 
present work. However, in case of nonlinear $NN$ folded potentials from RMF (NR3Y(NL3), NR3Y(TM1)), 
one need not require to add spin-orbit term from outside, as it is contained within the RMF\cite{sahu}.  

Finally reaction cross-sections and astrophysical reaction rates are calculated in the Hauser-Feshbach 
formalism using the computer package TALYS1.2\cite{talys}. 
Besides the phenomenological OMP, TALYS also includes the semimicroscopic nucleon-nucleus spherical optical model calculation 
with JLM potential\cite{jlm,jlm2}.
Therefore, for the sake of completeness, we compared our result
with the results obtained from JLM potential.

\section{Results}

For simplicity, this section is divided in three subsections. In the 
first subsection, results from RMF calculations are given. We will concentrate on the reaction cross-sections 
and astrophysical S factors in the second subsection. Furthermore, results for reaction rates for astrophysically 
important nuclei are provided. 
The third part is devoted to the 
effects of different $NN$ potentials in this mass region.

\subsection{RMF calculations}

\begin{table*}
\caption{Calculated  binding energy per nucleon\cite{adi} and charge radii\cite{angel} of selected $p$ nuclei
compared with experimental values.\label{be}}
\begin{center}
\begin{tabular}{lcccccccc}\hline
&\multicolumn{4}{c}{B.E./A(MeV)}
&\multicolumn{4}{c}{$r_{ch}$(fm)} \\
{}&FSUGold&TM1&NL3&Exp&FSUGold&TM1&NL3&Exp\\\hline
$^{102}$Pd&{8.480}&{8.537}&8.572&8.580& 4.460&{4.476}&{4.483}&4.483 \\
$^{106}$Cd& 8.494&{8.518}&8.532&8.539 &4.525&{4.535}&{4.535}&4.538  \\
$^{108}$Cd& {8.498}&{8.529}&8.537&8.550 &4.537&{4.549}&{4.552}&4.558 \\
$^{113}$In&{8.507}&{8.461}&{8.523}& 8.523&4.480&{4.575}&{4.588}&4.601 \\
$^{112}$Sn&{8.514}&{8.520}&8.502&8.514&4.595&{4.598}&{4.594}&4.594 \\
$^{114}$Sn&{8.534}&{8.526}&8.490&8.523&4.636&{4.611}&{4.662}&4.610 \\
$^{115}$Sn&{8.530}&{8.527}&8.494&8.514&4.607&{4.611}&{4.617}&4.615\\
$^{120}$Te&{8.461}&{8.461}&8.460&8.477&4.682&{4.688}&{4.735}&4.704 \\
\hline
\end{tabular}
\end{center}

\end{table*}

\begin{figure}
\resizebox{!}{!}{\includegraphics{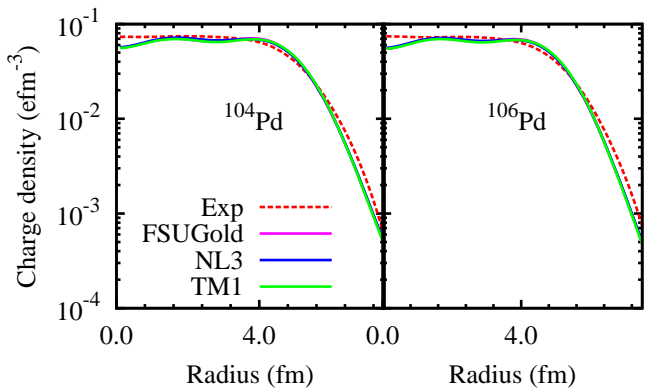}}
\caption{(Color online) Comparison of charge density from our calculation with  Fourier-
Bessel analysis of experimental electron scattering data\cite{de}
\label{cden}}
\end{figure}

In some earlier works \cite{gg,clah1,clah2,clah3,saumi}, FSUGold was proved to be successful in reproducing experimentally obtained 
binding energy, charge radius and charge density data in the A$\sim$55-100 region. Again in 1997, NL3 parameter set had been 
introduced by Lalazissis \etal\cite{lala} with a aim to provide a better description not only for the properties of stable nuclei 
but also for those far from the $\beta$ stability line and during last two decades, this parameter set successfully reproduces 
binding energy, charge radius etc of various elements throughout the periodic table\cite{lala,p2}.   
In order to confirm the applicability of RMF calculations in A$\sim$100-120 region, in Table \ref{be},
we compare nuclear binding 
energy per nucleon and charge radii of $p$ nuclei in the 
concerned mass region from our calculations with different sets of parameters with existing experimental data\cite{adi,angel}. 
We find that, in most cases, our calculations with
different sets of parameters 
match quite well with the experimental data.   
In figure \ref{cden} charge density from our calculations are compared with 
existing electron scattering data\cite{de} for Pd isotopes and here also, the agreement is well enough to confirm the credibility
of RMF models in this mass region.


\subsection{Astrophysical S factor and reaction rates}

In the present case, our calculations, being more
microscopic, are more restricting. In general, phenomenological models are usually fine tuned for nuclei
near the stability valley, but not very successful in describing elements near the proton and neutron rich regions.
Microscopic models, in contrary, can be extended to the drip line regions and 
therefore, this method can be used to study the reaction rates of nuclei involved in $p$ process 
nucleosynthesis network ($\sim$ 2000 nuclei are present in the total $p$ network). 
However, only a few number of stable $p$ nuclides are available in nature that can be accessed by the experiment and therefore
we are restricted to those nuclei for the purpose of comparison \footnote {One can apply this microscopic calculation
to study the neutron
capture reactions also, but at present, we are interested to study the proton capture reactions only as the desired $p$ network 
usually does not involve neutron capture reactions.}.

As a first test of the optical model potential, we have calculated elastic proton scattering at low energies where experimental
data are available. As the elastic scattering process
involves the same incoming and outgoing channel for the
optical model, therefore it is expected to provide the easiest way
to constrain various parameters involved in the calculation.
Here we are mainly interested in the energy region between 2-8 MeV as the
astrophysical important Gamow window lies within this energy range in the concerned mass region.
However, scattering experiments are very difficult at
such low energies, because the cross sections are extremely
small, and hence no experimental data are available. Therefore we have
compared the cross sections from our calculations with the lowest energy experimental data available in
the literature.

\begin{figure}
\resizebox{8cm}{7cm}{\includegraphics{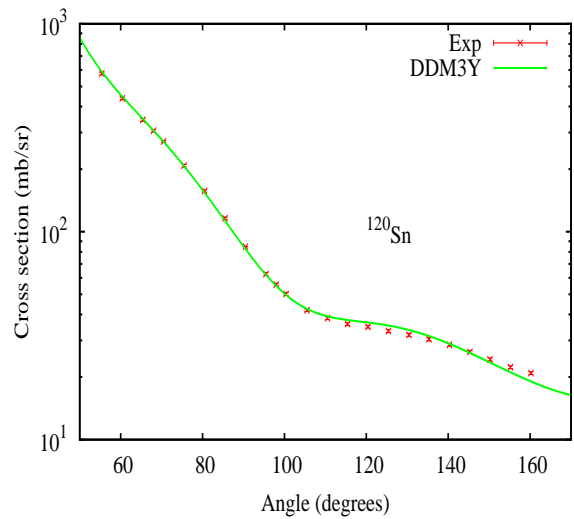}}
\caption{(Color online) Experimental and calculated cross sections for elastic
proton scattering at 9.7 MeV proton energy
 \label{120sn}}
\end{figure}

In figure \ref{120sn}, we present the result of our calculation with DDM3Y folded potential for $^{120}$Sn with 
available experimental data\cite{diff}.
To fit the experimental data, at first, the folded DDM3Y
potential is multiplied by factors 0.3 and 0.7 to obtain the
real and imaginary parts of the optical potential, respectively. 
However better fits for individual
reactions can be possible by varying different parameters. But if
the present calculation has to be extended to an unknown
mass region, this approach is clearly inadequate. Therefore,
we have refrained from fitting individual reactions. 
For example, in a previous work\cite{clah2}, the real and imaginary part of the potential was multiplied to 0.7 and 0.1 respectively.
But beyond that, the same set of parameters was unable to fit the experimental data for $p$ nuclei
in mass 80 region, and therefore the real and the imaginary part normalizations were chosen to be 0.81 and 0.15 respectively, 
in mass 90 region\cite{clah2}. However, there are no sharp boundaries for these mass regions, 
but for simplicity, we chose it in such a way that a single set of parameters can fit the entire mass region.
In this work, we consider A $\sim$ 100-120 region 
as we can trace the whole region with same set of parameters.

Yet, the astrophysical reaction rates depend on the proper choice of the level density and the E1 gamma strength. Therefore,
we have calculated all of our
results with microscopic level densities in Hartree-Fock (HF)
and Hartree-Fock-Bogoliubov (HFB) methods, calculated for
TALYS database by Goriley and Hilaire\cite{talys,gori} on the basis of Hartree-Fock calculations\cite{gori1}
. We have also
compared our results using phenomenological level densities
from a constant-temperature Fermi gas model, a back-shifted
Fermi gas model, and a generalized superfluid model from
TALYS. All these model parameters can be availed from TALYS database. 
We find that the cross sections are very sensitive to the
level density parameters, sometimes by a factor of 20\%.
We therefore analyzed, in most of the cases, the HF level densities fit the
experimental data better in this mass region.
Again, for E1 gamma strength functions, results derived from
HF + BCS and HFB calculations, available in the TALYS
database, are employed.  In this case also, the results for HF+BCS calculations describe the experimental data reasonably well 
and we present our results for that
approach only.

We now calculate some $(p,\gamma)$ cross sections relevant to $p$ nuclei in A$\sim$100-120 region where experimental
data are available.  At such low energies, 
reaction cross-section varies very rapidly making comparison between theory and
experiment rather difficult. Therefore the usual practice in low-energy
nuclear reaction is to compare another key observable, viz. the
S factor. It can be expressed as\cite{clah1}

\begin{equation}
S(E)=E\sigma(E)e^{2\pi\eta},
\end{equation}

where E is the energy in center of mass frame in keV which factorises out the pre-exponential low energy dependence of reaction
cross-section $\sigma(E)$, and $\eta$ indicates
 the Sommerfeld parameter with  
\begin{equation}
 2\pi\eta=31.29 Z_{p}Z_{t}\sqrt{\frac{\mu}{E}}.
\end{equation}

The factor exp$(2\pi\eta)$ is inversely proportional to the transmission 
probability through the Coulomb barrier with zero angular momentum(s-wave) 
and therefore removes exponential low energy dependence of $\sigma(E)$. 
Here $\sigma(E)$ is in  barn, $Z_{p}$ and $Z_{t}$ are the charge numbers 
of the projectile and the target, respectively and $\mu$ is the reduced 
mass (in amu) of the composite system. This S factor varies much slowly 
than reaction cross-sections and for this reason, we calculate this quantity
and compare it with experimentally obtained values.

\begin{figure}
\resizebox{8cm}{7cm}{\includegraphics{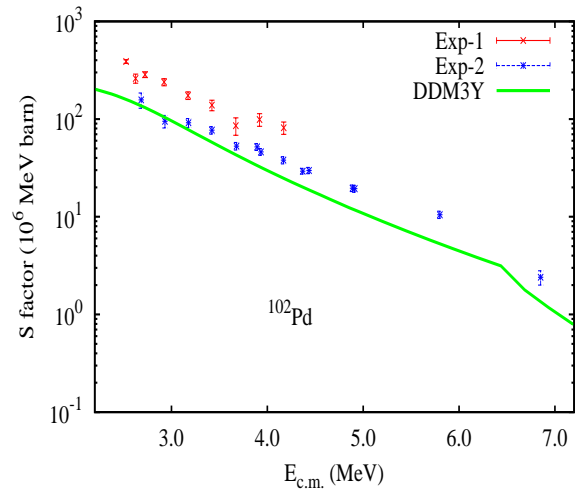}}
\caption {(Color online) S factors from two different microscopic potentials are compared with  experimental measurements
  for $^{102}$Pd. Here \textquotedblleft Exp-1\textquotedblright   is the experimental data from 
  reference\cite{pd102r1}, \textquotedblleft Exp-2\textquotedblright from reference\cite{pd102r2} and 
  \textquotedblleft DDM3Y\textquotedblright is for the DDM3Y-folded potential.\label{102pd}}
\end{figure}

\begin{figure}
\resizebox{8cm}{8cm}{\includegraphics{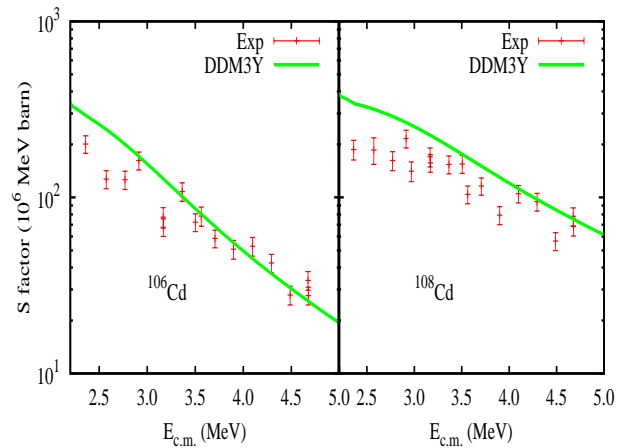}}
\caption {(Color online) S factors extracted from theory compared with experimental measurements
 for $^{106,108}$Cd. Here \textquotedblleft Exp\textquotedblright is the experimental data from reference\cite{cdref}. 
 \label{106cd}}
\end{figure}

\begin{figure}
\resizebox{8cm}{7cm}{\includegraphics{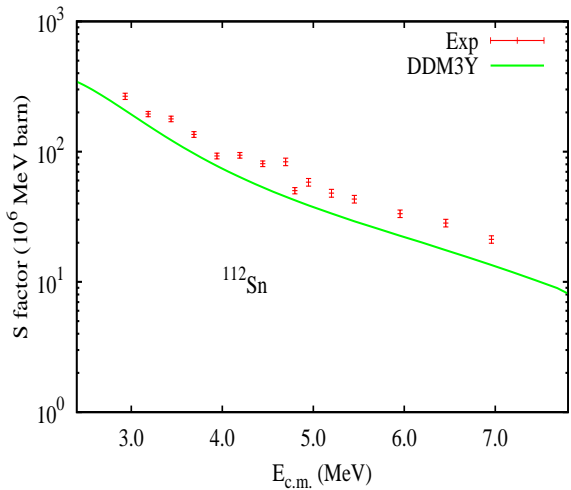}}
\caption{(Color online) S factors extracted from theory compared with experimental measurements
 for for $^{112}$Sn.  \label{112sn}}
\end{figure}

In figures \ref{102pd}-\ref{112sn} we present the results of some 
of our calculations with folded DDM3Y potential for Pd,Cd and Sn isotopes, respectively,
along with the corresponding experimental results. 
The experimental values for $^{102}$Pd are from reference\cite{pd102r1}(red point) and \cite{pd102r2}(green cross), 
$^{106,108}$Cd from Gy. Gy\"urky \etal\cite{cdref} and $^{112}$Sn from reference \cite{sn112ref}.

In case of $^{102}$Pd in figure \ref{102pd}, theoretical prediction 
is in a good agreement, mainly in the low energy regime, 
with the experimental data from reference \cite{pd102r2} but under 
predicts the data obtained from the reference\cite{pd102r1}. In reference \cite{pd102r1}, an activation technique was used 
in which gamma
rays from decays of the reaction products were detected off-line by two hyper-pure germanium
detectors in a low background environment, whereas in reference \cite{pd102r2}, cross-section measurements have been carried out at
the cyclotron and Van de Graaff accelerator by irradiation of thin sample
layers and subsequent counting of the induced activity. However, we can not comment on the individual merits of these experiments.

In case of $^{106,108}$Cd in figure \ref{106cd}, one can find that the agreement of theory with experimental
values are good enough, however there is a slight over prediction of $^{108}$Cd in the low energy regime. In case of $^{112}$Sn 
in figure \ref{112sn}, our calculation follows the experiment in a fairly good fashion.    

\begin{figure*}
\hspace{-2.5cm}
\resizebox{!}{!}{\includegraphics{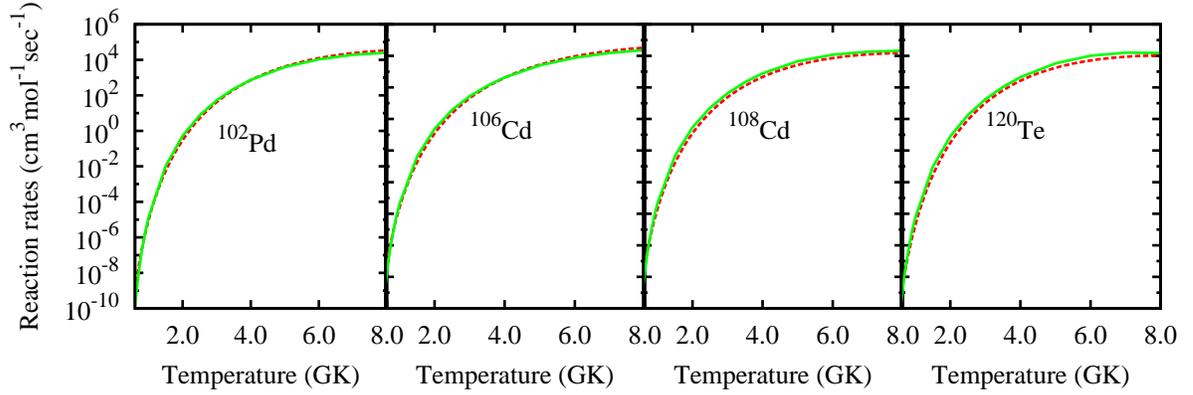}}
\caption {(Color online) Astrophysical reaction rates for $(p,\gamma)$ reactions of some important $p$ nuclei 
compared with NONSMOKER rates\cite{rate}. Here Green Continuous line: Present calculation, Red Dotted line: NONSMOKER calculation.
 \label{pgrate}}
\end{figure*}

\begin{figure*}
\hspace{-2.5cm}
\resizebox{!}{!}{\includegraphics{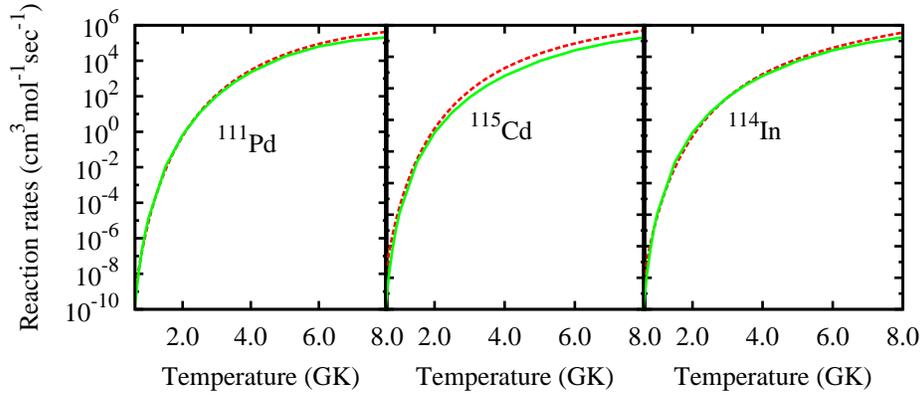}}
\caption {(Color online) Astrophysical reaction rates for $(p,n)$ reactions
compared with NONSMOKER rates\cite{rate}. Here Green Continuous line: Present calculation, Red Dotted line: NONSMOKER calculation.
 \label{pnrate}}
\end{figure*}

The success of this microscopic optical potential (DDM3Y interaction folded with FSUGold density)in reproducing S factor data for
the above $p$ nuclei leads us to calculate reaction rates
of some astrophysically important reactions. In figure \ref{pgrate}, we compare  
$(p,\gamma)$ reaction rates for some important $p$ nuclei 
 with NONSMOKER rates\cite{rate}. Again in figure \ref{pnrate}, reaction rates for charge exchange reactions $(p,n)$ for some nuclei,
 however not astrophysically significant enough, in this mass region are compared with existing NONSMOKER calculations. One
can see that the present calculation is very similar to the NONSMOKER values in almost all cases. 
Therefore, it is expected that all the results can
also be reproduced with commonly used NONSMOKER rates.

\begin{figure}
\resizebox{8cm}{7cm}{\includegraphics{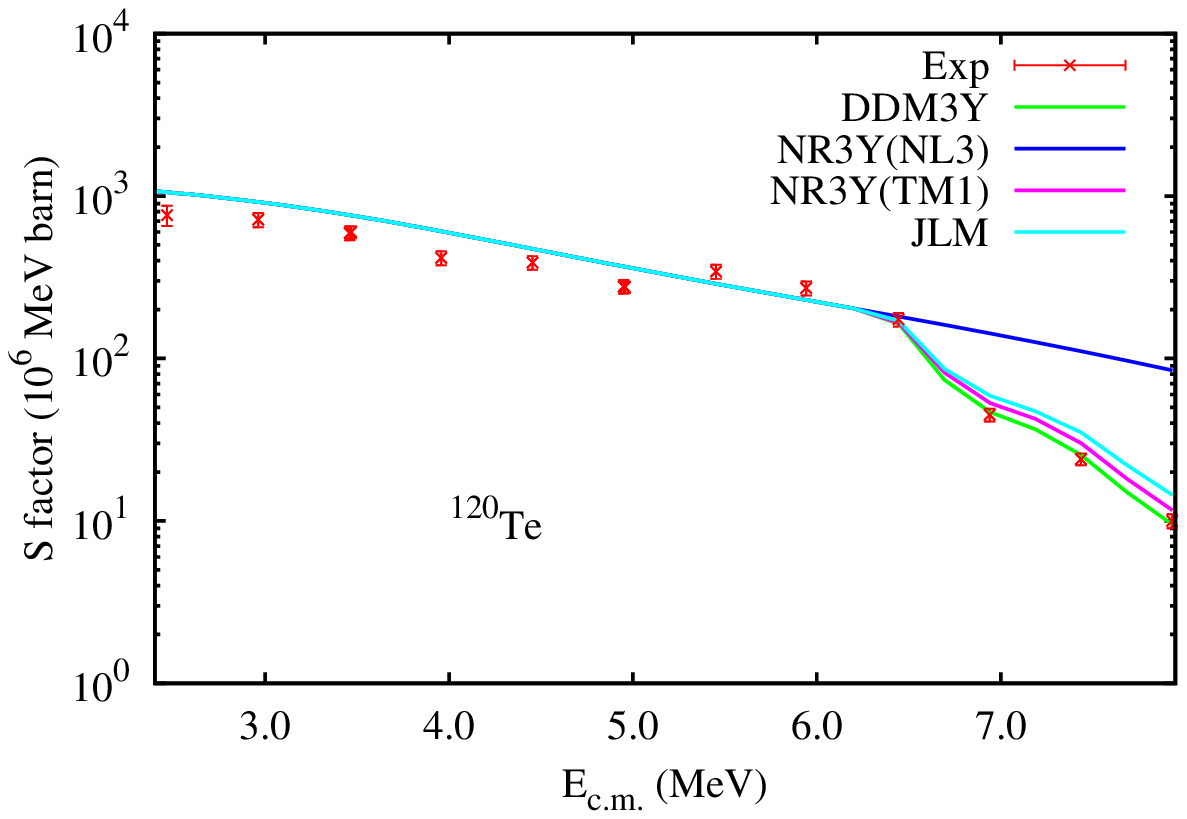}}
\caption {(Color online) S factors extracted from our calculations compared with experimental measurements
 for $^{120}$Te. Here \textquotedblleft Exp\textquotedblright is the experimental data from reference\cite{te120ref}. 
 For other details, see the text. \label{120te}}
\end{figure}

In the rest of the present manuscript, we mainly concentrate on the effects of optical potentials obtained by folding nonlinear
interactions from RMF.
In figure \ref{120te}, S factors for $^{120}$Te  obtained from NR3Y(NL3) and
NR3Y(TM1) potentials are compared with the experimental data taken from  reference\cite{te120ref}.
The S- factor with  DDM3Y interaction folded with  FSUGold density is also given 
for comparison. For better understanding, we have associated the result from JLM\cite{jlm, jlm2} potential also, 
which can be found on TALYS code.   
 
 One can see that our calculation with 
 folded DDM3Y potential shows a very nice agreement with experimental values throughout the energy range. The result
 from JLM interaction also agrees with the experiment. 
 In contrary, in case of 
 NR3Y(NL3)folded potential, there is a wide deviation of the theory with experimental data after 6 MeV whereas 
 the TM1 folded potential NR3Y(TM1) shows a decrease in S factor value around 6 MeV energy unlike the NR3Y(NL3)case.  
 
 The rapid drop of S factor
 values with increasing energy actually takes place due to the increasing contribution of higher angular momentum channels (l$>$0).
 Therefore, if the center of mass energy E$_{c.m.}$ becomes larger than the Coulomb barrier for a specific set of nucleon-nucleus
 reaction (E$_{c.m.}>$E$_c$), as a result the S factor will decrease rapidly with the growth of energy (E$_{c.m.}$)\cite{sf}.
 In the next subsection,
 we illustrate this physics in detail and show how this phenomenon is associated with different form of potentials.

\subsection{Optical potentials and effects for nonlinearity}

\begin{figure}
\resizebox{8cm}{7cm}{\includegraphics{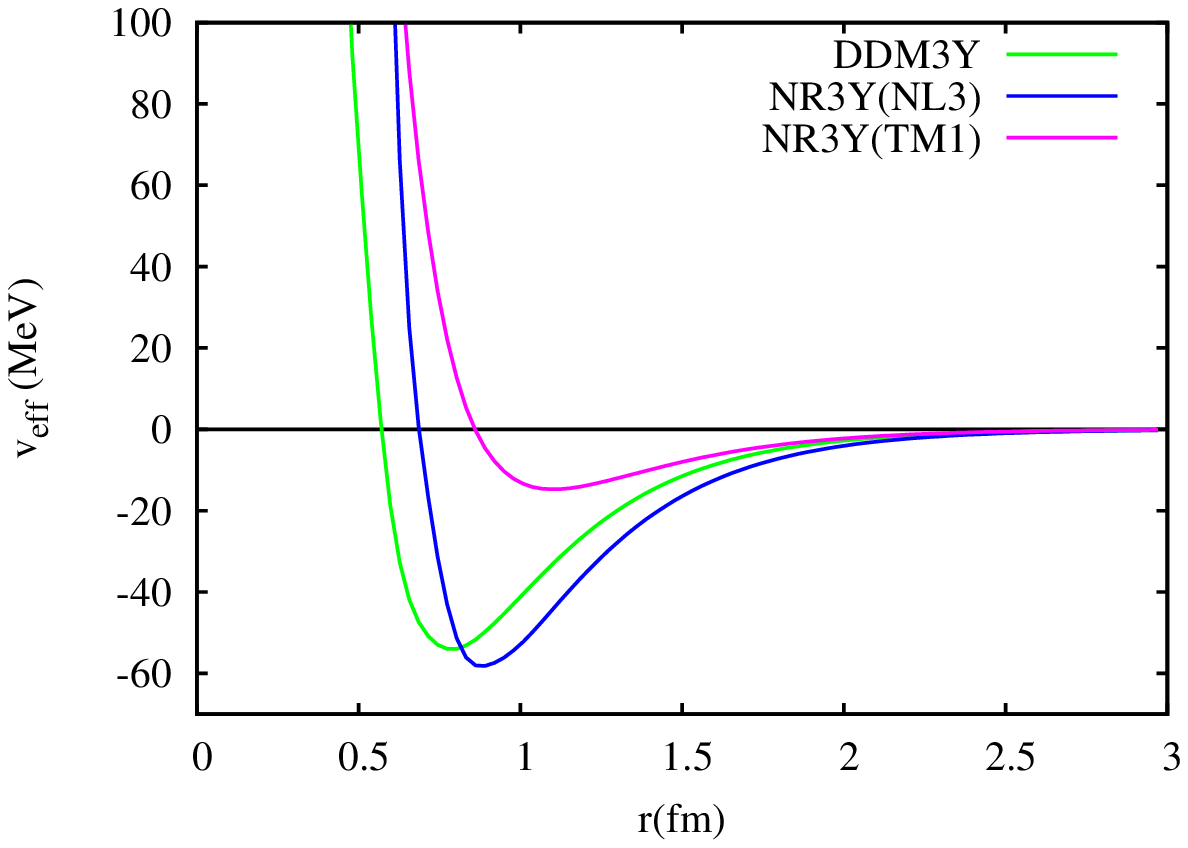}}
\caption{(Color online) Effective interaction potential for $^{120}Te$.
 \label{int}}
\end{figure}

We now interpret the above results (for example, see figure \ref{120te}) 
with the microscopic potentials obtained from different $NN$ interactions. In figure \ref{int}, 
the effective $NN$ interaction potentials (in MeV) are plotted with the radius r(fm) for $^{120}$Te. The DDM3Y interaction,
being dependent on the density, 
is different for different elements of the periodic table, whereas in contrary, other interactions remain unaltered for different
elements. In figure \ref{int} different forms of $NN$ interactions are given.
We find that the curves from DDM3Y and NR3Y(NL3) interactions generated from two different formalisms show almost similar
trend which
makes us believe that this nonlinear form of the $NN$ interaction 
can also be
used to obtain the microscopic optical potential. 

\begin{figure}
\resizebox{8cm}{7cm}{\includegraphics{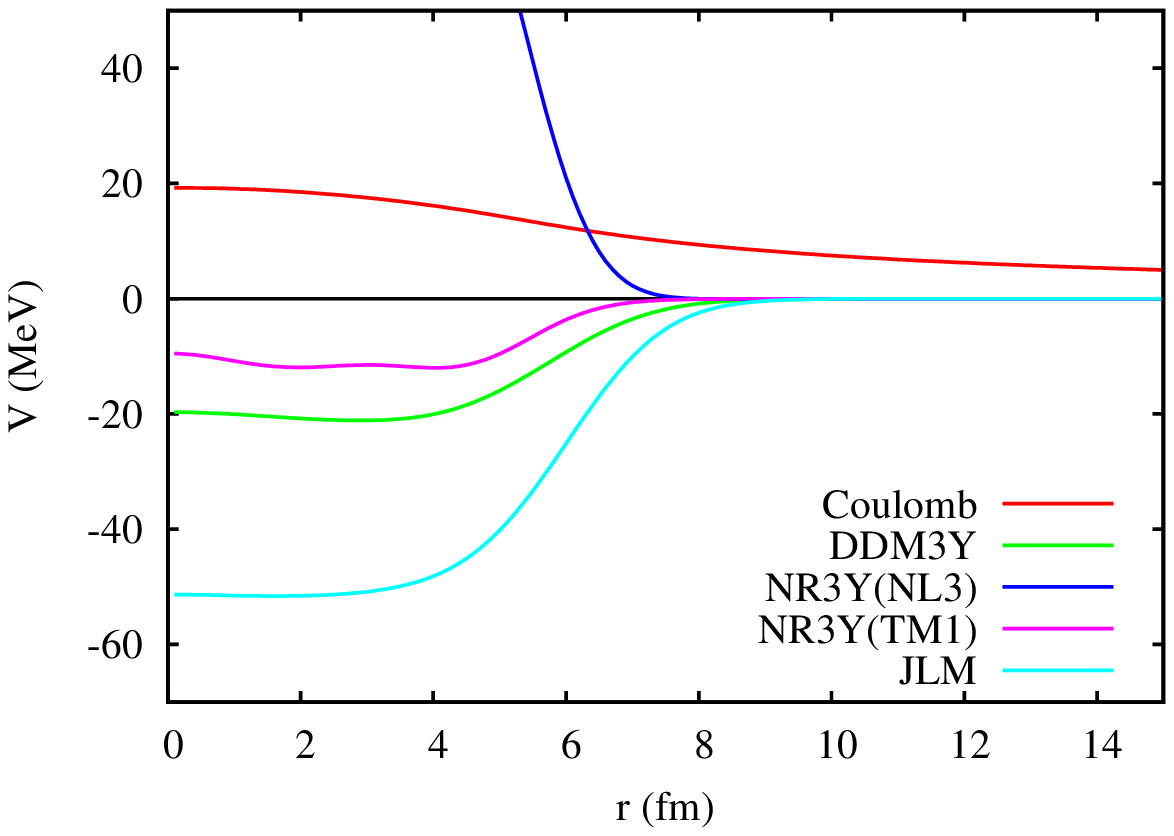}}
\caption{(Color online) Real central part of folded potentials and Coulomb potential for 7 MeV proton(Lab) incident on $^{120}$Te \label{pot}}
\end{figure}

A graphical representation of microscopic potentials for  $^{120}$Te after folding the interactions is represented
in figure \ref{pot}. Here the real central part of the optical potential is plotted with the radius. 
In the figure, one can see that the DDM3Y folded potential provides an attractive potential similar to the real part of the JLM
potential, whereas in
case of NR3Y(NL3) folded potential, the repulsive part overpowers the attractive part, as well as the Coulomb part of the potential. 
As a result, the resultant repulsive barrier becomes greater than the Coulomb barrier almost upto 
a range for a nuclear reaction to occur. Therefore the
penetrability of the higher angular momentum channels get reduced and as a obvious consequence, the desirable sharp drop in S factor 
(figure \ref{120te}) values has not been achieved. In case of TM1 folded potential, we can see that the effective contribution
of the optical potential is attractive in nature similar to the DDM3Y and JLM potentials and therefore, the Coulomb energy serves as 
the only repulsive barrier. As a result the 
penetration probability for higher angular momentum channels becomes higher than that of the NR3Y(NL3)case. This reason is replicated
as a drop of S factor
values at higher energies in figure \ref{120te}. In case of the imaginary part of the potential, the curves follow exactly the similar trend 
as that of the real part, i.e., apart from NR3Y(NL3) potential, rest of them gives attractive contribution. 
 One can explain the above scenario
from the numerical value of the nonlinear coupling
constant g$_3$(TM1 parameter set),
as given in Table \ref{nl3}, which is much less than that of the NL3 parameter set. Therefore it can be understood that with 
 decreasing values of the nonlinear coupling constants g$_2$ and g$_3$,
the repulsive
component of the optical potential also gets reduced and one point is attained when only the effect of 
Coulomb barrier remains as a dominating repulsive contributor 
and we will get patterns like JLM, TM1,
DDM3Y as shown in figure \ref{pot} and we can find the expected drop of S factor values at higher
energies due to the opening of 
 higher angular momentum 
channels. So from the above observations we can comment that there should be an 
upper cut-off for the coupling constant values of the nonlinear components.   

\section{Summary and Discussion}
To summarize, cross section for low energy $(p,\gamma)$ reactions for 
a number of $p$ nuclei in A$\sim$100-120 region have been calculated 
using microscopic optical model potential with the Hauser Feshbach reaction code TALYS. Mainly, microscopic potential 
is obtained by folding DDM3Y interaction with densities from RMF approach. Astrophysical reaction rates for $(p,\gamma)$ and $(p,n)$
reactions are compared with standard NONSMOKER results. Finally, the effect of microscopic optical potential obtained by folding 
nonlinear NR3Y(NL3) and NR3Y(TM1) interactions with corresponding RMF densities are employed to fit the experimental S factor data
for $^{120}$Te. The reason of the deviation of theoretical 
prediction with nonlinear NR3Y(NL3) potential from experiment at higher 
energies has been discussed and finally we made a comment on magnitude of 
the coupling terms of the nonlinear components that an upper cut-off value
 for $g_2$ and $g_3$ should be fixed to get proper repulsive component of 
the $NN$ interaction.

\end{document}